\documentclass{jfm}

\usepackage{graphicx}
\usepackage{epsfig}
\usepackage{bm}
\usepackage{amsmath}
\usepackage{amssymb}
\usepackage{wasysym}
\usepackage{epstopdf}
\usepackage{color}
\usepackage{xcolor}
\usepackage{epstopdf}
\usepackage{upgreek}
\usepackage{natbib}
\usepackage{arydshln}
\usepackage[T1]{fontenc}

\usepackage{natbib}

\usepackage{psfrag}
\usepackage{color}
\usepackage{graphicx,float,amssymb}

\usepackage{color}
\usepackage{graphicx}
\usepackage{natbib}
\usepackage{bm}
\usepackage{amsmath}

\usepackage[colorlinks=false, bookmarks=true]{hyperref}

\usepackage{subfigure}
\usepackage{graphicx}
\usepackage{epstopdf, epsfig}
\usepackage{graphicx}
\usepackage{epsfig}
\usepackage{bm}
\usepackage{amsmath}
\usepackage{amssymb}
\usepackage{wasysym}
\usepackage{epstopdf}
\usepackage{color}
\usepackage{xcolor}
\usepackage{epstopdf}
\usepackage{upgreek}
\usepackage{natbib}
\usepackage{arydshln}
\usepackage[T1]{fontenc}

\usepackage{natbib}

\usepackage{psfrag}
\usepackage{color}
\usepackage{graphicx,float,amssymb}

\usepackage{color}
\usepackage{graphicx}
\usepackage{natbib}
\usepackage{bm}
\usepackage{amsmath}

\usepackage{graphicx}
\usepackage{epsfig}
\usepackage{bm}
\usepackage{amsmath}
\usepackage{amssymb}
\usepackage{wasysym}
\usepackage{epstopdf}
\usepackage{color}
\usepackage{xcolor}
\usepackage{epstopdf}
\usepackage{upgreek}
\usepackage{natbib}
\usepackage{arydshln}
\usepackage[T1]{fontenc}

\usepackage{graphicx}
\usepackage{epsfig}
\usepackage{bm}
\usepackage{amsmath}
\usepackage{amssymb}
\usepackage{wasysym}
\usepackage{epstopdf}
\usepackage{color}
\usepackage{xcolor}
\usepackage{epstopdf}
\usepackage{upgreek}
\usepackage{natbib}
\usepackage{arydshln}
\usepackage[T1]{fontenc}

\shortauthor{Mathai, Loeffen, Chan, and Wildeman}

\title{Dynamics of heavy and buoyant underwater pendulums}

\author{Varghese Mathai\aff{1,2}$\dagger$, 
	Laura A. W. M. Loeffen\aff{2}\footnote{Equally contributed authors ; ~ ~$\ddagger$ swildeman@gmail.com},
	Timothy~T.~K.~Chan\aff{2,3},
	and Sander Wildeman\aff{4,2}$\ddagger$ }

\affiliation{\aff{1} School of Engineering, Brown University, Providence, RI 02912, USA.
	
	\aff{2} Physics of Fluids Group and Max Planck Center for Complex Fluids,\\ Faculty of Science and Technology,
	University of Twente, P.O. Box 217, 7500 AE Enschede, The Netherlands.
	
	\aff{3} Department of Physics, The Chinese University of Hong Kong, Shatin, Hong Kong.
	
	\aff{4} Institut Langevin, ESPCI, CNRS, PSL Research University, 1 rue Jussieu, 75005 Paris, France.
}

\begin{document}
	
	\maketitle

	\begin{abstract} 
	The humble pendulum is often invoked as the archetype of a simple, gravity driven, oscillator. Under ideal circumstances, the oscillation frequency of the pendulum is independent of its mass and swing amplitude. However, in most real-world situations, the dynamics of pendulums is not quite so simple, particularly with additional interactions between the pendulum and a surrounding fluid.  Here we extend the realm of pendulum studies to include large amplitude oscillations of heavy and buoyant pendulums in a fluid. We performed experiments with massive and hollow cylindrical pendulums in water, and constructed a simple model that takes the buoyancy, added mass, fluid (nonlinear) drag, and bearing friction into account. To first order, the model predicts the oscillation frequencies, peak decelerations and damping rate well.  An interesting effect of the nonlinear drag captured well by the model is that for heavy pendulums, the damping time shows a non-monotonic dependence on pendulum mass, reaching a minimum when the pendulum mass density is nearly twice that of the fluid. Small deviations from the model's predictions are seen, particularly in the second and subsequent maxima of oscillations. Using Time- Resolved Particle Image Velocimetry (TR-PIV), we reveal that these deviations likely arise due to the disturbed flow created by the pendulum at earlier times. \textcolor{black}{The mean wake velocity obtained from PIV is used to model an extra drag term due to incoming wake flow. The revised model significantly improves the predictions for the second and subsequent oscillations.}
	
	\end{abstract}
	

	\vspace{-1.5 cm}
	\section{Introduction}

{The first known study on pendulums dates back to Galileo Galilei in 1605. The story goes that Galileo observed the lamplighter pushing one of the swaying chandeliers in the Pisa cathedral. Galileo timed the swings with his pulse and concluded that, although the amplitude decreased, the time of each swing remained constant. 
\textcolor{black}{Half a century later this observation inspired Christiaan Huygens to invent the pendulum clock, which until 1930 has set the standard for accurate timekeeping. Based on Newton's laws of motion Huygens could derive that an ideal, frictionless pendulum has a period $T= 2  \pi \sqrt{L/g}$ for small amplitudes, where $L$ is its length, and $g$ the acceleration due to gravity~\citep{huygens1986christiaan}. However, as is well known to any clockmaker, real pendulums are often far from ideal. Even in vacuum, and when friction is negligible, one has to take into account the finite amplitude and  mass distribution of the pendulum in order to correctly predict its period. When a pendulum swings in a fluid such as air or water, additional fluid forces have to be taken into account. In fact, it was through careful observations of deviations from ideal pendulum motion that the nature and strength of many of these fluid forces were brought to light. Concepts like buoyancy, added mass and fluid friction (viscosity) emerged concurrently with the study of pendulums swinging in different fluid environments~\citep{stokes1851effect}.}}


\textcolor{black}{In recent years there has been a renewed interest in the use of simple pendulums to probe fluid-structure interactions.}
\cite{neill2007pendulum} conducted experiments on steel and brass pendulums oscillating in different fluids. Their results agreed fairly well with the motion predicted by taking the buoyancy and added mass into account, implying that viscous corrections played only a minor role in the oscillation frequency. 
In an extension to this, \cite{bolster2010oscillating} conducted experiments on a  pendulum oscillating in a viscous fluid.
In all of the above studies, the researchers focused on small amplitude vibrations of dense pendulums in viscous fluids, where the dynamics could be reasonably modelled using the ideal flow approximation, with the inclusion of a linear drag in the equation of pendulum motion.  {\color{black}Many studies have focused on the flow disturbances induced by an oscillating body in a fluid. For one-dimensional oscillations at low to moderate Reynolds numbers, the effect of a body's oscillation on the wake around it has been studied in some detail~\citep{tatsuno1990visual,tatsuno1993wavy}. Various types of wakes have been quantified depending on the body oscillation frequency, velocity, fluid viscosity and the characteristic length scale of the problem.} \textcolor{black}{More recent studies have addressed the influence of flow disturbances on pendula stability. \cite{obligado2013bi} found that the equilibrium of a pendulum facing an incoming flow displays both bi-stability and hysteresis. Further, flow disturbances due to incoming turbulence can modify the stability map by changing the drag acting on the pendulum. }




\begin{figure} 
	\centering
	\includegraphics[width= 0.85\textwidth]{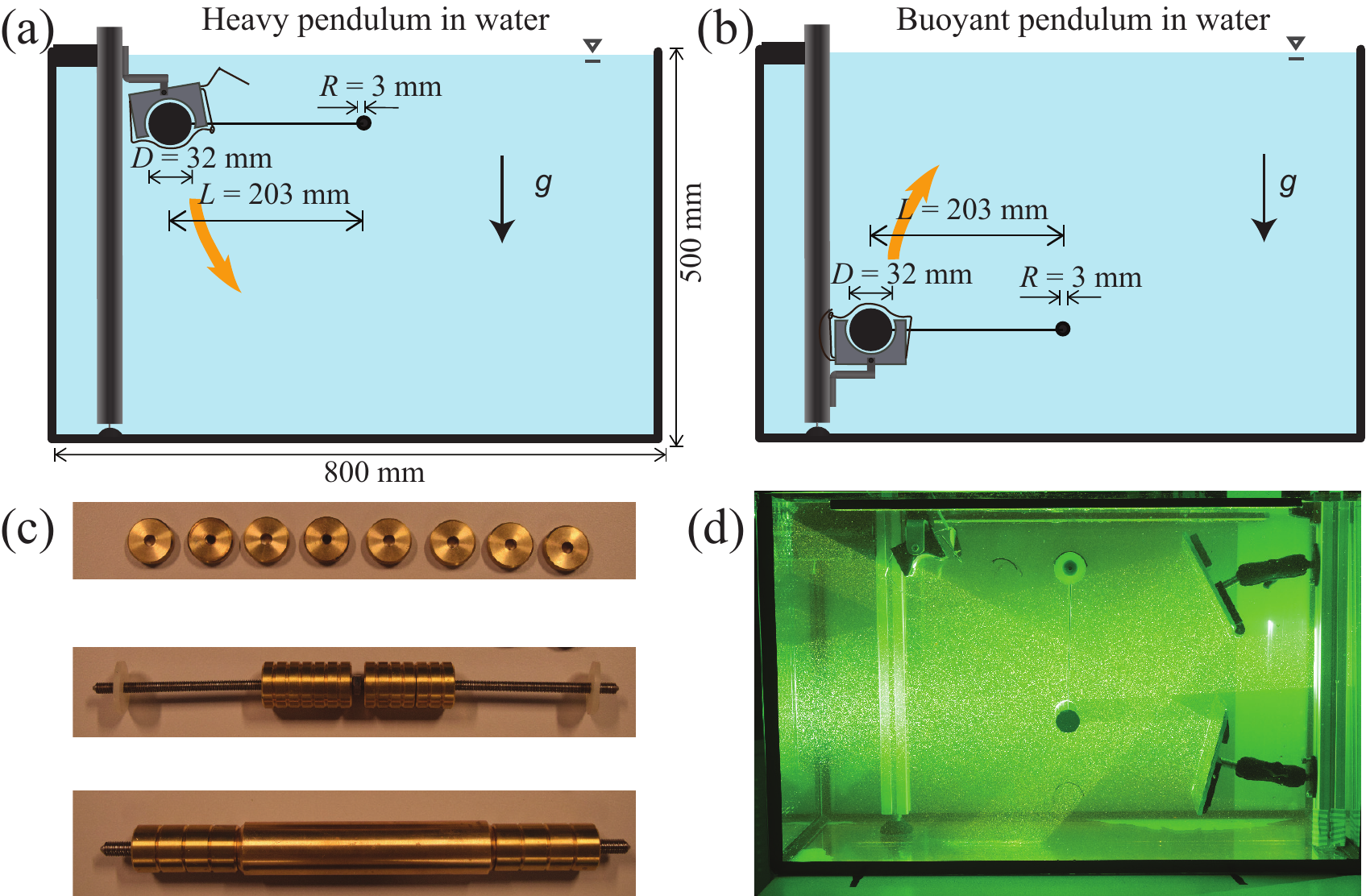}
	\caption{Schematics of the experimental setup for (a) heavy and (b) buoyant pendulums. (c)~Cylinder mass was varied by inserting brass disks into the hollow cylinder. (d) Experimental setup used for performing PIV measurements.}
	\label{setup}
	\vspace{-.4 cm}
\end{figure}

\textcolor{black}{Here we extend the scope of pendulum-fluid interaction studies to the regime of large amplitude motions of  heavy and buoyant cylindrical pendulums in a dense fluid.}  
\textcolor{black}{We have performed controlled experiments in which we simultaneously track the motion of the pendulum and the flow field around it, in a regime where the Reynolds number (based on the cylinder diameter) Re $\sim \mathcal{O}(10^4)$.} The control parameter in our experiment is the cylinder to fluid density ratio $m^* \equiv \rho_p/\rho_f$ The ratio of the cylinder diameter to the pendulum length $D/L$, the fluid properties~($\rho_f$ and $\nu$) and the release angle~($\theta_0 = 90^\circ$) are fixed. 
We vary $m^*$, covering both buoyant~($m^* < 1$) and heavy~($m^* > 1$) pendulums. We build on the knowledge of buoyancy, added mass and inertial drag to construct a basic model for the pendulum motion, which we compare in detail to the experiments. \textcolor{black}{We find that a reduced added mass as compared to the ideal flow case leads to better agreement between experiment and model predictions. We use insights from particle image velocimetry (PIV) measurements to model a higher-order fluid-structure interaction, by which initially induced fluid motions affect the force on the pendulum at later instants.}

%

\vspace{-.4 cm}

\section{Experiments}

The experiments were conducted using a cylindrical pendulum, with length $L = 203$~mm, diameter $D = 32$ mm and span $W = 300$ mm. This resulted in a cylinder span to diameter ratio $W/D \approx 10$. The cylinder attached to the pendulum arm was constructed from polyvinyl chloride~(PVC) pipe sealed with two plastic caps. Masses could be inserted inside the pipe~(see Fig.~\ref{setup}(c)), which let us study both heavy and buoyant pendulums in water.

The pendulum was placed in a water tank (800 mm $\times$ 400 mm $\times$ 500 mm), and pivoted on a metal rod (radius $R = 3$ mm), which was fixed to the walls using suction cups. \textcolor{black}{The wire connecting the cylinder to the pivot axis had a diameter of 1.2 mm. This was thick enough to not bend during oscillations, while also providing low resistance to the flow.} The cylinder ends remained at least $1.3D$ away from the side walls. The lowest position of the heavy cylinder during its motion was $5D$ above the base of the water tank, and similarly, the highest position of the buoyant cylinder during its motion was $5D$ below the water level. Schematics of the basic experimental arrangement for  the heavy and buoyant cases are shown in Fig.~\ref{setup}(a) \& (b), respectively. \newline 
The release mechanism consisted of a semicircular cavity with two curved metal wire segments. The cylinder was held against the semicircular cavity, and locked in place by the metal wires. The release was done by pulling the wire, which ensured a gentle release of the pendulum without creating disturbances in the water. By inverting these parts, the buoyant pendulums could also be released without disturbing the water. The release position was at  a $90^\circ$ angle from the vertical. A waiting interval of 15 minutes ensured that any residual flow from a previous run had damped out.  

A high-speed camera~(Photron Fastcam 1024PCI) was used to record the pendulum motion. The cylinder movement was detected using a circle detection method~\citep{mathai2015wake}. \textcolor{black}{The standard deviation of error in centre detection was around 0.27 mm for all cases, i.e. $<$ 1\% of the pendulum diameter.} Additionally, we performed Time-Resolved Particle Image Velocimetry~(TR-PIV) to measure the flow field surrounding the pendulum. The flow was seeded with fluorescent polystyrene tracer particles (diameter $\approx 125 \ \mu$m). The seeding particle size was small enough to have both a small Stokes number and a small Stokes/Froude ratio~\citep{mathai2016microbubbles}. A high-speed Laser~(Litron LDY-303HE) along with cylindrical optics was used to create a light sheet through the mid-section of the cylinder. Mirrors were placed as shown in Fig.~\ref{setup}(d), which ensured that the shadows cast by the cylinder were removed by reflected light. This enabled us to have well-lit tracer particles all around the cylinder. \textcolor{black}{A double-frame camera~(Imager sCMOS) was used at a frame rate of 25 fps. The particle seeding density for PIV was such that an interrogation window contained around 6 particles, which is considered good for accuracy. The inter-pulse time $\Delta t$ was varied from 1--3 ms. $\Delta t$ was optimised by ensuring that most of the tracer particles remained within the interrogation window, with a maximum movement $\approx 1/4^{th}$ of the window width. The PIV analysis was done with two stage processing and 50 percent overlap using LaVision software.} 

\vspace{-.4 cm}

\section{Model equation of motion}

\begin{figure}
	\centering
	\includegraphics[width=.9\textwidth]{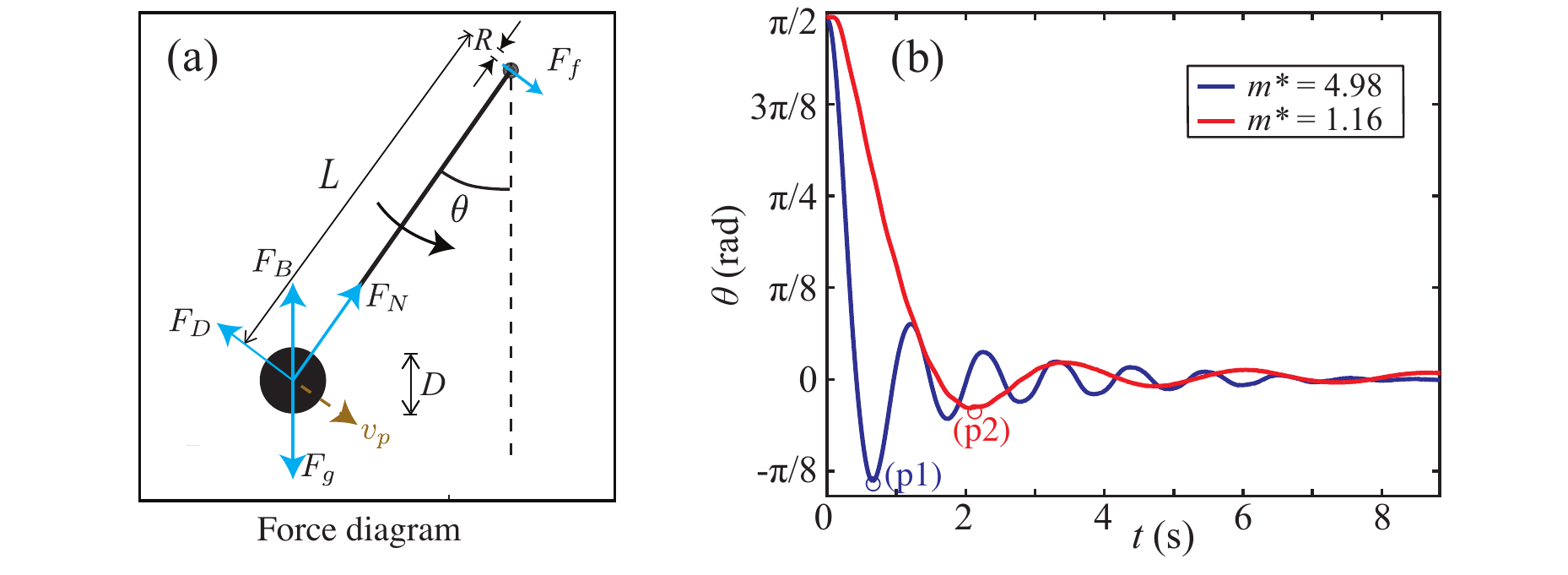}
	\caption{(a) Schematic of the cylindrical pendulum at an instant when it swings to the right. Blue arrows indicates the forces on the pendulum: buoyancy ${F_B}$, weight ${F_g}$, drag ${F_D}$, normal force due to the pendulum arm ${F_N}$, and a bearing friction ${F_f}$. The dashed arrow and $v_p$ denote the instantaneous velocity of the cylinder. (b) Angular position ($\theta$) vs time for $m^* = 4.98$ and $m^* = 1.16$ cases. (p1) \& (p2) -- peak deceleration points.}
	\label{PendulumSwing}
		\vspace{-.4 cm}
\end{figure}

The equation of motion of a pendulum oscillating in a fluid can be expressed by Newton's law in the angular form:
\begin{align}
I \frac{d^2 \theta}{dt^2} &= \sum{\tau},
\label{eqn:Equation of Motion}
\end{align}
Here, $\theta$ is the angular position of the pendulum with respect to the equilibrium position, $I$ the moment of inertia of the system and ${\tau}$ the net torque acting on the system~(see Fig.~\ref{PendulumSwing}(a)). The pendulum mass and moment of inertia come mostly from the cylinder at a distance $L$ from the rotation axis. Since $2 L/D > 10$,  the cylinder moment of inertia about its mass centre $I_{cm} = m D^2/8$ is negligible compared to the pendulum moment of inertia $m L^2$. 
The effect of the pendulum accelerating through the fluid is modelled by means of an added mass $m_a$. Therefore, the effective mass of the system can be written as $m_{eff} = m + m_a$. With these approximations, we model the system as a point mass with a moment of inertia: $I_p= m_{eff} L^2$.

The torques acting on the system can be written as $\sum {\tau} = \sum {r} \times {F}$, where ${F}$ are the individual forces, and ${r}$ the corresponding arm lengths. The forces acting on the system are gravity $F_g \equiv \rho_p \mathcal{V} g$, buoyancy $F_B \equiv \rho_f \mathcal{V} g$, fluid drag $F_D$ and a bearing friction $F_f \approx \mu_f F_N$ (see Fig.~\ref{PendulumSwing}(a)). Here, $\rho_p$ and $\rho_f$ are the particle and fluid densities, respectively, $\mathcal{V}$ is the cylinder volume, $\mu_f$ is the friction coefficient and $F_N \approx (F_B - F_g) \cos \theta$ is the normal force acting at the bearing. \textcolor{black}{The tension due to the centrifugal force is negligible.} {The Galileo number $\text{Ga} \equiv \sqrt{g D^3 |m^*-1|}/\nu$. The velocity scale in Ga is a gravitational  velocity $v_g = \sqrt{gD |1-m^*|}$. The Reynolds number is defined as Re~$ = v_{max} D/\nu$, where $v_{max}$ is the measured maximum velocity of the cylinder in the first swing. We note that Ga~$\sim~\mathcal{O}(10^3 - 10^4$) gives a predictive estimate of the Re range in the experiment~(see Table~\ref{ReGa}).} 
The drag force may be written as $F_D  = \frac{1}{2}\rho_f A_p C_D \ v_p^2$, where $C_D$ is the drag coefficient, $A_p$ is the projected area of the cylinder and $v_p \equiv  L d\theta /dt$ is the instantaneous cylinder velocity~\citep{batchelor2000introduction}. \textcolor{black}{By defining $m^* \equiv \rho_p/\rho_f$ and $\tilde{t} \equiv t\sqrt{g/L}$, the equation of motion can be written in non-dimensional form as}

\begin{align}
m_{eff}^* \frac{d^2 \theta}{d\tilde{t}^2}= - k \sin{\theta} \ - \ c \left| \frac{d \theta}{d \tilde{t}}\right| \frac{d \theta}{d \tilde{t}} \ - \ h \left | \cos{\theta} \right | \ \text{sgn} \left( \frac{d \theta} {d\tilde{t}} \right)
\label{EqtofMotion},
\end{align}
where $m_{eff}^* = (m^* + m^*_a); \ k = \left |m^* - 1 \right |; \ c = \frac{1}{2} C_D \frac{A_p L}{\mathcal{V}};$  and $h = \mu_f\left| m^* -1 \right| \frac{R}{L}$.\\

	\begin{table}
		\caption{Ga, and Re for a selected number of mass density ratios $m^*$ used in the experiment.}
		
		\centering
		\begin{tabular}
			{p{0.14\linewidth}p{0.06\linewidth}p{0.04\linewidth}p{0.04\linewidth}p{0.04\linewidth}p{0.04\linewidth}p{0.04\linewidth}p{0.04\linewidth}p{0.04\linewidth}p{0.04\linewidth}p{0.04\linewidth}p{0.04\linewidth}}
			
			\hline
			\hline
			~ & \vline~~ & ~ & Heavy~ & ~ & ~ & ~ &  & \vline~ & Buoyant & ~ & ~          \\
			
			\hline
			
			$m^*$ & \vline~4.98 & 4.15 & 3.65 & 2.82 & 2.20 & 1.79 & 1.37 & \vline~0.95 & 0.75 & 0.54 & 0.33          \\
			\hline Ga $\times 10^{-3}$ & \vline~35.8 & 31.8 & 29.2 & 24.2 & 19.6 & 15.8 & 10.9 &\vline~3.8 & 9.0 & 12.2 & 14.6         \\
			\hline			Re $\times 10^{-3}$  &\vline~33.7 & 30.9 & 28.8 & 24.7 & 20.6 & 16.9 & 11.9 &\vline~4.2 & 10.0 & 13.7 & 16.6          \\  
			
		\end{tabular}
		\label{ReGa}
		\vspace{-.05 cm}
	\end{table}

In order to solve eq~\ref{EqtofMotion}, $m_a^*$, $C_D$ and $\mu_f$ have to be set. \textcolor{black}{It is well known that added mass coefficients $m_a^*$ for oscillating cylinders depend on a variety of parameters, including the frequency of oscillation,  distance to boundaries, free surfaces, etc. Most studies have focused on the case of cylinders near a free surface and for small oscillations~\citep{dong1978effective,konstantinidis2013added,koo2015simplified,tatsuno1990visual,mathai2017mass,mathai2018flutter}. However, in complex situations with relative motions, curved trajectories and unsteady three-dimensional wakes with flow separation, $m_a^*$  can deviate significantly from the two-dimensional potential flow added mass coefficient. Moreover, the fact that the cylinder span is finite will induce a three-dimensional (3-D) flow near the cylinder ends, allowing some fluid to move to the sides. 
Therefore the cylinder motion is expected to accelerate less fluid in the 3-D case as compared to 2-D potential flow. Nevertheless, to begin with, we choose the 2-D potential flow added mass coefficient of an infinite cylinder, $m_a^* =1.0$.}

The drag coefficient range in our experiments could be estimated from  the range of Re~(see table~\ref{ReGa}). 
\textcolor{black}{{We begin with a simplified mean drag coefficient $C_D \approx 1.2$ based on the Reynolds number range in our experiments~\citep{lienhard1966synopsis}. Note that the influence of Reynolds number and vortex shedding on $C_D$ are not included at this stage, and will be discussed later.} $\mu_f$ is expected to lie in the range [$0.2 -0.3$] for lubricated steel-on-steel contact. We estimated $\mu_f \approx 0.2$, based on tests performed using very heavy pendulums in air, since in these cases the drag and added mass forces were at a minimum.}

	Fig.~\ref{PendulumSwing}(b) shows $\theta$ vs $t$ predicted by the model for heavy pendulums with $m^* = 1.16$ and $m^* =4.98$.
	Clearly, the amplitude decay and oscillation frequency depend on the mass density ratio. With this in mind, we explore the predictions for a range of $m^*$, covering  heavy and buoyant underwater  pendulums.

%

\section{Results}

\subsection{Model predictions}

\begin{figure}
	\centering
	\includegraphics[width=0.99\textwidth]{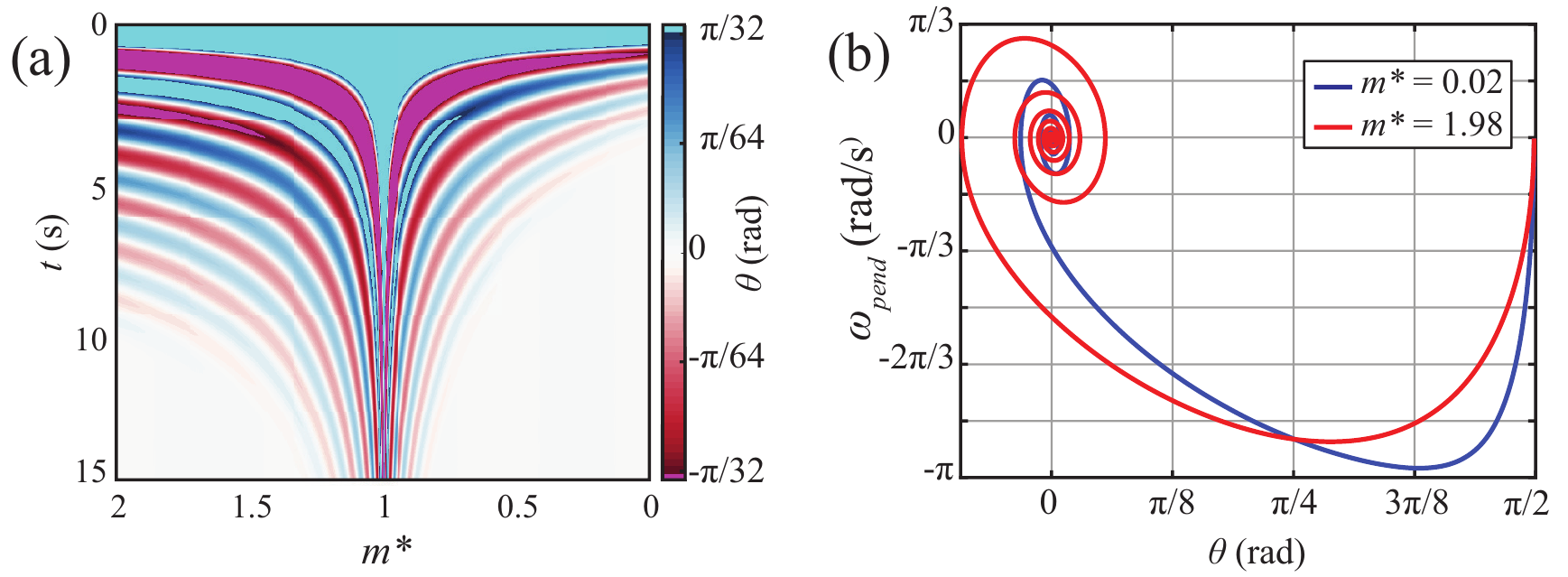}
	\caption{(a) Contour-plot of $\theta$ vs time $t$ and mass-density ratio $m^*$. (b) Phase portraits from numerical solution of eq.~\ref{EqtofMotion} showing the evolution of angular velocity $\omega_{pend}$ vs angular position $\theta$ for {\color{black}$m^* = 0.02$ and $m^* = 1.98$ cases.} Note that these cases were chosen  with identical driving: $|F_B - F_g|$. Phase portraits obtained for available experimental data~(not shown here) show similar behaviour.}
	\label{ModelResults}
	\vspace{0 cm}
\end{figure}

{\color{black}We first present some typical pendulum motions predicted by the model equation of motion.}
\textcolor{black}{In Fig.~\ref{ModelResults}(a), we show a contour plot of the angular position $\theta$ vs time $t$  for continuous variation of $m^*$. We notice a clear asymmetry about $m^* = 1$ line, which separates the heavy cases from the buoyant cases. 
For instance, the oscillation frequency is slightly higher for a buoyant cylinder as compared to a heavy cylinder that has identical driving $|{F_B} - {F_g}|$. On the buoyant side ($m^* <1$), the oscillations damp faster with decreasing $m^*$, owing to the decreasing inertia. Other aspects of the pendulum motion are altered in going from $m^*>1$ to $m^*<1$. In Fig.~\ref{ModelResults}(b), we show the phase portraits of two cases, one buoyant~{\color{black}($m^* = 0.02$)} and other heavy~{\color{black}($m^* = 1.98$)}, with identical driving $|{F_B} - {F_g}|$.  {\color{black}The low mass-density ratio of $m^* = 0.02$ chosen here corresponds to a cylinder made from a very light material such as expanded polystyrene~(\cite{mathai2015wake}).} The buoyant pendulum with ($m^* = 0.02$) accelerates quickly, reaching a high angular speed in a short period of time. At the same time, owing to its low inertia, which in this case comes entirely from the fluid~(or $m_a^*$), the motion damps out quickly. Therefore, the peak angular displacement at first minimum $\theta_0$ is lower for the buoyant pendulum as compared to the heavy pendulum~($m^* = 1.98$), while the peak angular velocity reached $\omega_{peak}$ is higher for the buoyant cylinder.} 

%

\subsection{Model vs experiment}

\begin{figure}
	\centering
	\includegraphics[width=0.93\textwidth]{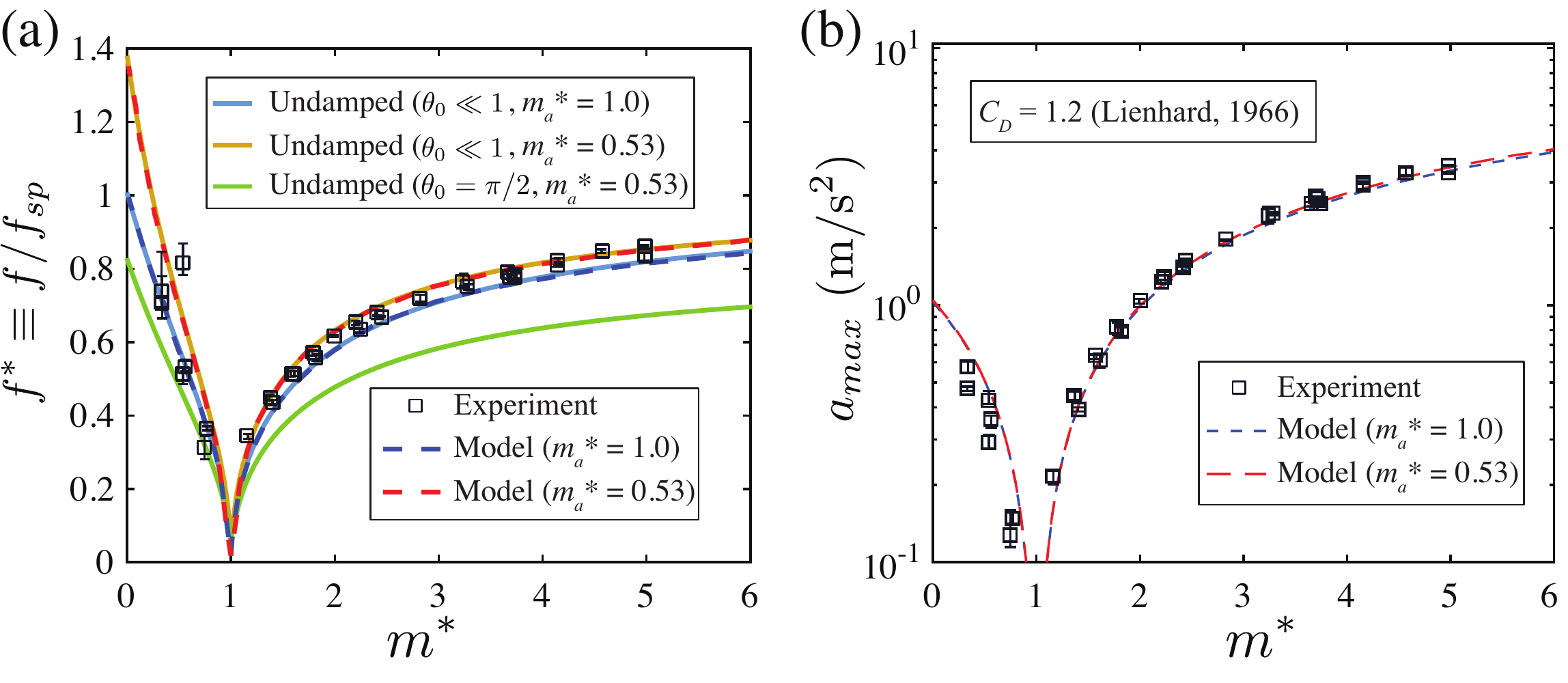}
	\caption{ (a) Normalized oscillation frequency $f^*$ from experiment and model~(eq.~\ref{EqtofMotion}). Here, $f$ is averaged over the first four oscillations, and normalised by $f_{sp} = \frac{1}{2 \pi} \sqrt{\frac{g}{L}}$. {\color{black} For the full equation of motion, using potential flow added mass $m^*_a = 1.0$ leads to an underprediction of $f^*$. For a reduced $m^*_a = 0.53$, the predictions of eq.~\ref{EqtofMotion} match well with the experiments.  Remarkably, an undamped model for small amplitude simple pendulum, and with the same $m^*_a = 0.53$, reproduces the curve, providing evidence that the nonlinear drag plays only a weak role in the oscillation frequency $f^*$. The green curve shows that an undamped model with large initial amplitude $\theta_0 = \pi/2$ underpredicts the oscillation frequency.} (b) Peak pendulum deceleration, $a_{max}$ vs $m^*$.}
	\label{f&a}
	\vspace{-.20 cm}
\end{figure}

\textcolor{black}{We now present a one-to-one comparison between the model and experiment.  In Fig.~\ref{f&a}(a), we show the  normalised frequency of oscillation. As expected, the frequency of oscillation has a clear dependence on the mass ratio $m^*$. When the pendulum mass density is close to the fluid density~(or $m^* \to 1$), the frequencies are low. For heavy cylinders, the frequency increases with $m^*$, and asymptotically approaches the frequency of a large amplitude simple pendulum $f^*_{\pi/2}$. The predictions of the model~(eq.~\ref{EqtofMotion}) are given by the blue and red dashed curves.} 
The blue curve corresponds to the prediction when the potential flow added mass $m_a^* = 1.0$ is used. This slightly underpredicts the oscillation frequency. A best fit to the data is obtained for $m_a^* = 0.53$. Thus, for $C_D =1.2$, $m_a^* =0.53$ and $\mu_f = 0.2$, the  frequency predictions of the model are in good agreement with the experiments for both buoyant and heavy cylinders.

It is interesting to compare the predictions of the model~(Eq.~\ref{EqtofMotion}) when the drag and the bearing friction are ignored altogether. The solid green curve in Fig.~\ref{f&a}(a) shows the frequency prediction for large amplitude oscillations when the potential flow added mass alone is accounted for. This underpredicts the oscillation frequency. However, when the same undamped model is used for small amplitude oscillations, the predictions improve significantly~(see solid blue curve). This is because the nonlinear drag is highly effective in quickly reducing the oscillation amplitudes to modest values. A further improvement is obtained upon using a reduced added mass of $m^*_a = 0.53$. \cite{neill2007pendulum} had noted that linear drag had only a small effect on the oscillation frequency.  Here we found that the same holds for the nonlinear drag term.


The frequency of oscillation represents an averaged quantity for the pendulum motion considered here. A  more sensitive quantity would be the peak deceleration $a_{max}$ of the oscillating pendulum, which signifies the instant when the largest force acts on the pendulum. 
Following the initial instants of release of the cylinder, a peak in the deceleration is experienced when the cylinder reaches the end of the first swing, i.e. points such as (p1) \& (p2) in Fig.~\ref{PendulumSwing}(b).  In Fig.~\ref{f&a}(b), we compare the experimentally measured peak deceleration with the predictions of the model~(eq.~\ref{EqtofMotion}). The peak deceleration changes significantly with $m^*$, and the model predictions are in reasonable agreement with the experimental measurements. Interestingly, the value of the added mass coefficient has only a minor role in $a_{max}$. Instead, the fluid drag has a leading role in the peak deceleration of the pendulum.
Thus, the frequency curve~(Fig.~\ref{f&a}(a)) provides a probe that singles out the added mass effect, while the deceleration curve~(Fig.~\ref{f&a}(b)) probes mainly the nonlinear drag's effect.

Figure~\ref{ModelResults}(a) also provides insight into an interesting aspect of the oscillation decay for heavy pendulums. For $m^* > 1$ cases, the oscillations decay quicker with increasing $m^*$. This occurred despite the higher inertia of heavier pendulums, and seems counter-intuitive based on our common knowledge of pendulums oscillating in air.
However, this trend cannot continue to large $m^*$, since in the limit of very large inertia ($m^* \gg 1$) the decay rate should reduce with $m^*$. This implies the existence of an optimal heavy pendulum ($m_{\text{opt}}^*>1$) for which one observes the quickest damping.  We will provide an approximate model to predict $m^*_{\text{opt}}$, corresponding to the quickest damped pendulum.

\begin{figure}
	\centering
	\includegraphics[width=0.95\textwidth]{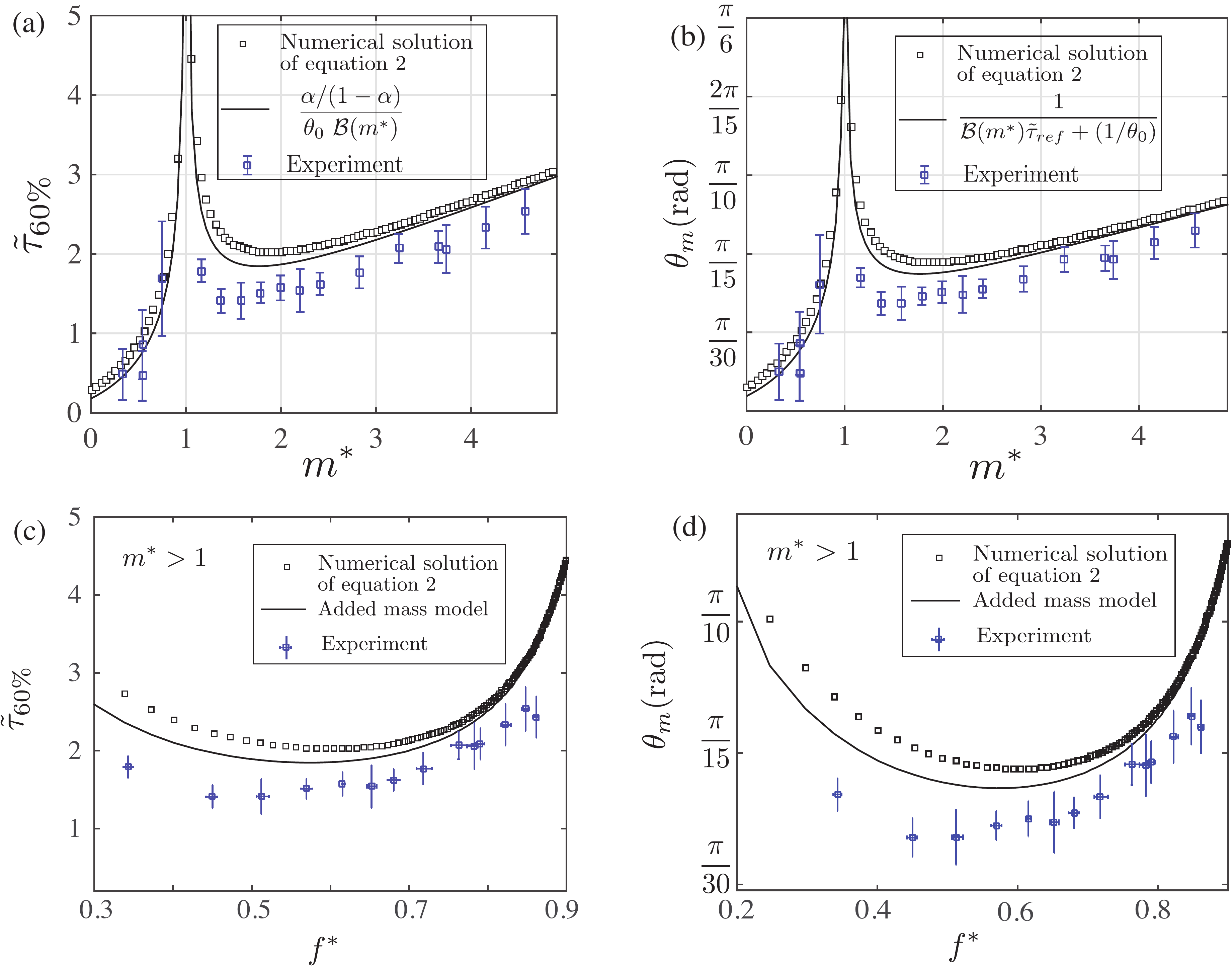}
	\caption{\textcolor{black}{(a) Time to decay 60\% of the initial amplitude $\tilde{\tau}_{60\%}$ vs mass density ratio $m^*$. (b)~Amplitude envelope $\theta_m$ vs $m^*$ after a time $\tau_{ref} = 3 \pi \sqrt{L/g}$. For the heavy pendulums, an optimal $m^*$ and optimal $f^*$ are visible in both experiment and simulations. (c) $\tilde{\tau}_{60\%} $ vs normalised oscillation frequency $f^* = f/ f_{sp}$, where $f_{sp} = \frac{1}{2 \pi} \sqrt{\frac{g}{L}}$. (d) $\theta_m$ vs $f^*$ at $\tau_{ref} = 3 \pi \sqrt{L/g}$.  Note that (c) \& (d) show heavy cases only. The basic model overpredicts the decay time and amplitude envelope for all cases.} }
	\label{tau_60_theta_3by2T_L}
	\vspace{-.2 cm}
\end{figure}

\subsection{Amplitude decay}

{\color{black}
	The initial damping is determined mainly by the nonlinear drag; therefore, we ignore the bearing friction force in the equation of motion (eq.~\ref{EqtofMotion}). This yields: $m_{eff}^* \frac{d^2 \theta}{d\tilde{t}^2} = - k \sin{\theta} - c \left| \frac{d \theta}{d \tilde{t}}\right| \frac{d \theta}{d \tilde{t}}$. In the absence of fluid drag~($c = 0$), the solution is given by $\theta (\tilde{t}) = \theta_0 \cos{\omega_0 \tilde{t}}$, with $\omega_0 = \sqrt{k/m_{eff}^*}$, and $\theta_0$ the initial angle. If the damping constant $c$ is small such that the pendulum amplitude is approximately constant over one period, we can assume that the solution is still approximately the same, but with a maximum amplitude $\theta_m \big{(}\tilde{t}\big{)}$ which slowly decays in time. Consequently, the energy in the oscillations evolves as $E\big{(}\tilde{t}\big{)} \approx \frac{1}{2} k \ \theta_m \big{(}\tilde{t}\big{)}^2$.
	The overall decay rate of the energy can be equated to the average work done by the drag force over one period $T = 2\pi /\omega_0$: 
	
	\begin{align}
	\frac{d E}{d \tilde{t}} = \frac{\omega_0}{2 \pi} \int_{0}^{\frac{2 \pi}{\omega_0}} F_D\big{(}\tilde{t}\big{)} \dot{\theta} \big{(}\tilde{t}\big{)} d\tilde{t}
	\label{energydamp}
	\end{align}
	
	Using $E\big{(}\tilde{t}\big{)} \approx \frac{1}{2} k \ \theta_m \big{(}\tilde{t}\big{)}^2$ and $F_D = - c \ \dot \theta^2$, we obtain a differential equation for the slowly varying envelope $\theta_m (t)$ of the form: $\frac{d\theta_m}{d\tilde{t}} \approx - \mathcal{B} (m^*)  \theta_m^2$, where $\mathcal{B} (m^*) = \frac{4 c}{3 \pi} \sqrt{\frac{|m^* -1|}{(m^* +m_a^*)^3}}$.  This yields the time to decay to a certain fraction ($1- \alpha$) of the initial amplitude $\theta_0$ as $\tilde{\tau}_{(1-\alpha)} = \frac{\alpha/(1-\alpha)}{\theta_0 \ \mathcal{B}(m^*) }.$

	%
For $m^* > 1$, and using the best fit $m_a^* =0.53$, $\mathcal{B}$ has a maxima at $m_{\text{opt}}^* \approx 1.75$. In other words, for a given $\theta_0$ the damping is the quickest when the pendulum is nearly twice the density of the fluid. \textcolor{black}{Alternately, one could plot the amplitude envelope $\theta_m$ after a dimensionless time $\tilde{{\tau}}_{ref} \equiv \frac{\tau_{ref}}{\sqrt{L/g}}$ for different $m^*$. This yields an expression $\theta_m = \frac{1}{\mathcal{B}(m) \tilde{\tau}_{ref}+  (1/\theta_0)}$. In Fig.~\ref{tau_60_theta_3by2T_L}(a)--(b) we show comparisons between the model and experiment for $\tilde{\tau}_{60\%}$ and $\theta_m$, respectively, vs mass-density ratio $m^*$.  The black symbols denote the model predictions by solving the full equation of motion~(eq.~\ref{EqtofMotion}). The black curve shows the prediction by the approximate model described above. Note that we first fit an envelope to the amplitude decay, using the maxima~(peaks) of $\theta$ vs $t$ curve, i.e a curve that grazes through the maxima of the amplitude curve. The optimal damping at $m^*_{\text{opt}} \approx 1.75$ is clearly visible in both the simulations and the reduced model. The blue data points are from experiment. A similar behaviour with a minima at an optimal $m^*$ is visible in the experiments. At the same time we note that both $\tilde{\tau}_{60\%}$ and $\theta_m$ are lower in the experiment. Further, the optimal $m^*$ value in experiment is slightly lower as compared to the model predictions.}  For very heavy pendulums ($m^* \gg 1$), the damping $\mathcal{B} \propto 1/m^*$, which leads to the linear relation $\tilde{\tau}_{60\%} \propto m^*$. This is also the commonly encountered situation for pendulums oscillating in air.

\textcolor{black}{The above discussed non-monotonic damping behaviour may be understood as the interplay between the higher speeds achieved by the heavier pendulums, and the nonlinear growth of the drag in proportion to the square of the speed.}
	\textcolor{black}{The phenomenon observed here may hold some analogies to the added damping for oscillating cylinders~\citep{dong1978effective}, which is usually expressed in terms of the oscillation frequency. Therefore, in Fig.~\ref{tau_60_theta_3by2T_L}(c)--(d)  we plot $\tilde{\tau}_{60\%}$ and $\theta_m$, respectively, as a function of the normalised oscillation frequency $f^*$. The $\theta_m$ plot shows a clear minima at $f^* \approx 0.6$. Similar to the $m^*$  plots~(Fig.~\ref{tau_60_theta_3by2T_L}(a)--(b)), the model overpredicts both $\tilde{\tau}_{60\%}$ and $\theta_m$, which occurs despite using the optimal added mass coefficient $m_a^* = 0.53$ determined earlier. Clearly, the origin of these must lie in the inadequacy of the simplified drag model used~($C_D =1.2$). Nevertheless, the constant $C_D$ case can be viewed as the most basic model, which captures many essential features of the damping. Using a higher value of $C_D$ seems the logical choice. In the following we will show that accounting for a few phenomenological effects can improve the predictions. We will introduce these modifications to the model and equation of motion.}

{\color{black} \subsection{Drag corrections}}

\textcolor{black}{Firstly, the influence of the instantaneous Re on $C_D$ can be taken into account.} This leads to a faster decay at small amplitudes (or small $v_p$), since $C_D$ is higher at lower Reynolds numbers~\citep{hoerner1965fluid}. Secondly, the forces due to vortex shedding behind the cylinder could be modelled into the drag term. The drag coefficient in the presence of vortex shedding may be approximated as $C_{Dvs} = C_D (1 + k \sin{\omega_{vs} t} )$, where $k \sim 0.1$~\citep{govardhan2000modes}, and $\omega_{vs} = 2 \pi \ \text{St} \ v_p/D$ is the vortex shedding frequency. Here St is the Strouhal number~\citep{govardhan2005vortex}, $v_p$ the cylinder velocity and $D$ the cylinder diameter. This leads to a marginal reduction in the predicted peaks of the oscillations.

Fig.~\ref{PIV}(a)--(b) compares the experiment and model predictions for the angular position~$\theta$ vs time $t$ for a heavy pendulum with $m^*$ = 4.98. 
\textcolor{black}{The oscillation frequency matches well between the experiment and the model. 
However, deviations are seen in the second peak of oscillation, marked as {\it (r)} in Fig.~\ref{PIV}(b).  The deviation persists for all of the subsequent oscillations.} This suggests that the major factor causing the deviations between the model and experiment is still missing in the equation of motion.


\begin{figure} 
	\centering
	\vspace*{0 cm}
	\includegraphics[width=0.9\textwidth]{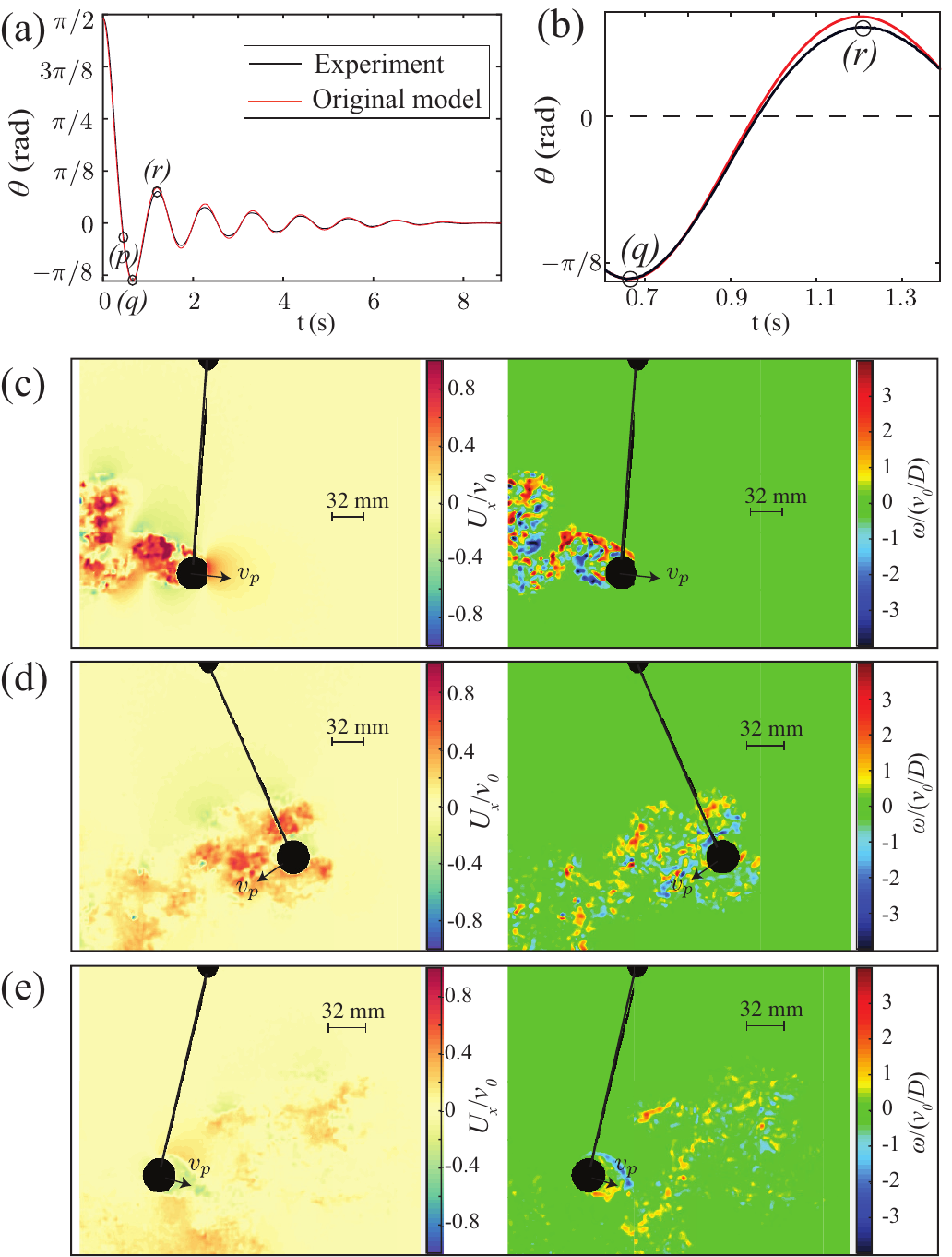}
	\caption{(a)Angular position $\theta$ vs time t for $m^* = 4.98$. The red curve shows the model predictions. (b) shows zoom-in of the second swing. The model overpredicts the maximum amplitude~(point marked as \textit{(r)} in (a) \& (b)).  (c) -- (e) Normalised horizontal velocity $U_x/v_0$~(\textit{left}) and normalised vorticity $\omega/(v_0/D)$~(\textit{right}) at the three instants marked in subfigure~(a) as \textit{(p)}, \textit{(q)}, and \textit{(r)}, respectively. Here, $v_0$ is the measured maximum speed of the pendulum.}
	\label{PIV}
	\vspace{-.0 cm}
\end{figure}

\subsection{PIV and wake corrections}
To understand the origin of the deviation in the second and subsequent peaks of the oscillations, we employed high-speed PIV to quantify the flow around the cylinder during its motion. 
Fig.~\ref{PIV}(c) shows the normalised velocity and vorticity fields at a time instant when the cylinder is moving towards the right during its first swing (point {\it(p)} in Fig.~\ref{PIV}(a)). The wake behind the cylinder is clearly visible, while the flow ahead of  the cylinder has negligible vorticity i.e. nearly irrotational~(Fig.~\ref{PIV}(c)(\textit{right})). The wake is unsteady and is expected to induce unsteady forces on the cylinder during its swing.

One can expect that the mean drag on the cylinder at this instant is fairly predicted  by the nonlinear drag term $F_D$ in the equation of motion~(eq.~\ref{EqtofMotion}). Consequently, the first swing of the pendulum is captured well by our model.
Next we focus at a later instant in time~(point {\it (q)} in Fig.~\ref{PIV}(a) \& (b)), when the cylinder has just completed its first swing and is returning to the left. Fig.~\ref{PIV}(d) shows the velocity and vorticity fields at this instant. In this case, however, the cylinder is moving towards a disturbed background flow. The horizontal velocity plot~(see \textit{left-side} plot of Fig.~\ref{PIV}(d)) indicates that the cylinder faces an incoming flow to the right. The effect of this disturbed flow is not taken into account in our model. Therefore, as the cylinder travels further through the disturbed flow, it experiences a greater resistance than what is predicted by the drag term in our model. Fig.~\ref{PIV}(e) shows the flow field at the instant when the cylinder completed its reverse swing~(point {\it (r)} in Fig.~\ref{PIV}(a) \& (b)). By this time, the original wake in front of the cylinder appears to have dissipated. 

\begin{figure}
	\centering
	\includegraphics[width=0.97\textwidth]{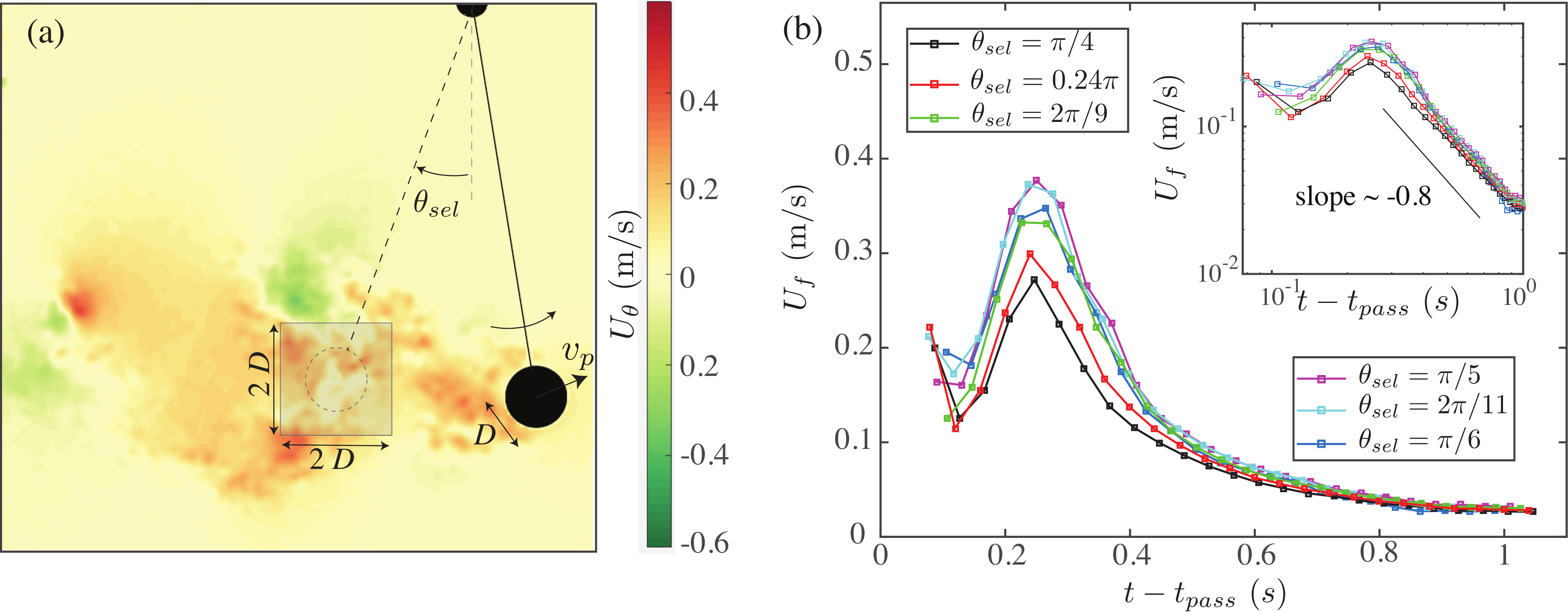}
	\caption{\textcolor{black}{(a) Schematic showing the estimation of mean wake velocity $U_f$ from a sample PIV flow field. A sample video of the PIV flow field is provided as supplemental material. $U_f$ is computed inside a $2D \times 2D$ window located at a selected angular position $\theta_{sel}$.  (b) Decay of wake velocity $U_f$ in time at various angular positions $\theta_{sel}$.  $t_{pass}$ is the time when the cylinder passed $\theta_{sel}$. The initial oscillation of the wake shows clear indication of vortex shedding. However, after this phase the wake decay is smooth. Inset shows the same plot on log-log scale. While at short times $U_f$ shows oscillations, the long-time decay shows a power law decay.}}
	\label{Wakedecay_time}
	\vspace{-.0cm}
\end{figure}

\textcolor{black}{One of the approximations of our model is that the cylinder velocity $v_p$ nearly equals the relative velocity $v_{rel}$ between the cylinder and the flow. However, the PIV snapshots reveal a strong mean flow after the first swing of the pendulum. This disturbed flow in the wake could result in $v_{rel}$ deviating significantly from $v_p$, which is the likely reason for the overprediction of the maxima of the second and subsequent oscillations. We look at the mean fluid velocity $U_f$ at different angular positions in the cylinder wake. Fig.~\ref{Wakedecay_time}(a) shows details of the $2D \times 2D$ window at a selected angular position $\theta_{sel}$ used in the estimation of $U_f$. The evolution of $U_f$ in time $t - t_{pass}$ is shown in Fig.~\ref{Wakedecay_time}(b). Here, $t_{pass}$ is the time when the cylinder passed $\theta_{sel}$.} 
\textcolor{black}{The wake decay shows some interesting characteristics. Initially, $U_f$ oscillates, which is a clear indication of vortex shedding in the cylinder wake during the initial moments after the cylinder has passed $\theta_{sel}$. We also see variations in the peak for different angular positions $\theta_{sel}$. This part of the curve is highly unsteady and difficult to model. However, later in time $U_f$ decays in a gradual and monotonic way. The inset to the figure shows the same plot on log-log scale. The long-time decay appears to follow a power law decay.}

\textcolor{black}{Several studies have addressed the decay of wakes for rising and falling particles~\citep{wu1993sphere,bagchi2004response}. These revealed many aspects of the temporal and spatial decays of wakes behind spheres in quiescent fluids, and in turbulent flows. \cite{almeras2017experimental} disentangled the wake decay behind rising bubbles into near wake and far fields, with their characteristic decay rates. The wake decay observed in Fig.~\ref{Wakedecay_time} also represents the combined effect of the local decay of the wake and the advection of the wake with a velocity $U_f$, i.e. $\frac{d{U_f}}{dt} = \frac{\partial{U_f}}{\partial t} + \frac{U_f}{L} \frac{\partial{U_f}}{\partial \theta}$. These terms can be very difficult to disentangle for the wide range of $m^*$~(or Re) in our experiments. Alternatively, one can look at the wake velocity at locations just upstream of the cylinder during its return swing. Fig.~\ref{Wakedecay_thetanorm}(a) shows the normalised wake velocity $\tilde{U}_f$ as a function of normalised angular position $\theta^*$ for four different mass ratios.  Here  $U_f$ is normalised by the pendulum speed at equilibrium position of each $m^*$ case, thus representing a characteristic speed for that $m^*$. For the heavier pendulums, $v_{p0}$ is comparable to the gravitational velocity scale $v_g$, but it deviates as $m^*$ is reduced. Similarly, the maximum angle reached by the pendulum at the end of its first swing $\theta_{max}$~(point $(q)$ in Fig.~\ref{PIV}(a)(b)) is used in the normalisation for $\theta$. For clarity, we also show the arrow of increasing time, indicating that the return swing starts at $\theta^* \to 1$. With these normalisations the data show a reasonable collapse. It is remarkable that the data collapse despite the large variation in $m^*$: [1.15, 4.98], which corresponds to a Reynolds number variation of almost a decade (Re $\in$ [4000, 34000]). At the initial instant, i.e. when $\theta^* \to 1$, the cylinder sees a large incoming flow velocity. As the pendulum swings back to $\theta^* \sim 0$, the wake velocity has almost completely decayed. Beyond this we observe a surprising increase in the incoming flow velocity. This increase arises from the cylinder wake during the first swing, where the pendulum had its highest swing velocity. Since the cylinder velocity at this position was large, the wake has not decayed completely.}

	 We obtain an approximate fit for the variation of $\tilde{U}_f$ vs $\theta^*$~(see Fig.~\ref{Wakedecay_thetanorm}(a)).  {\color{black} With the wake flow modelled using $\tilde{U}_f(\theta^*)$ shown above, we can include the effect of the incoming flow through a modified drag term: $ F_{Df} \approx \frac{1}{2} \rho_f A_p C_D \ (v_p - U_f)^2$, where $U_f = \tilde{U}_f \ v_{p0}$. This can be implemented as a modified torque  that replaces the original drag term in eq.~\ref{EqtofMotion}. The modified equation of motion reads: 
	
	\begin{align}
	m_{eff}^* \frac{d^2 \theta}{d\tilde{t}^2}= - k \sin{\theta} \ - \ c \left| \frac{d \theta_{rel}}{d \tilde{t}}\right| \frac{d \theta_{rel}}{d \tilde{t}} \ - \ h \left | \cos{\theta} \right | \ \text{sgn} \left( \frac{d \theta} {d\tilde{t}} \right)
	\label{EqtofMotion_mod},
	\end{align}
	where the relative angular velocity $ {d\theta_{rel}}/{d\tilde{t}} = {(v_p - U_f)}/{\sqrt{L g}}$. Note that Eq.~\ref{EqtofMotion_mod} requires no new input parameters, and works for the full range of Reynolds number in our experiments, i.e. $\text{Re} \in [4000, 34000]$.  The improvements in the predictions of the amplitude decay are presented in Fig.~\ref{Wakedecay_thetanorm}(b), which shows the error in the predicted amplitude of the second peak of oscillation for all mass ratios ($m^* >1$) studied.} The predictions of the revised model are shown along with those of the original model. Accounting for the wake flow results in significant improvement in the second peak for all $m^*$ cases. The mean error reduces from $\sim$ 14.8 \% to 3.8 \% after including the influence of the wake flow into the equation of motion. These results demonstrate the importance of fluid-structure coupling for pendulums oscillating at large amplitudes in viscous fluids.}


\begin{figure}
	\centering
	\includegraphics[width=1.0\textwidth]{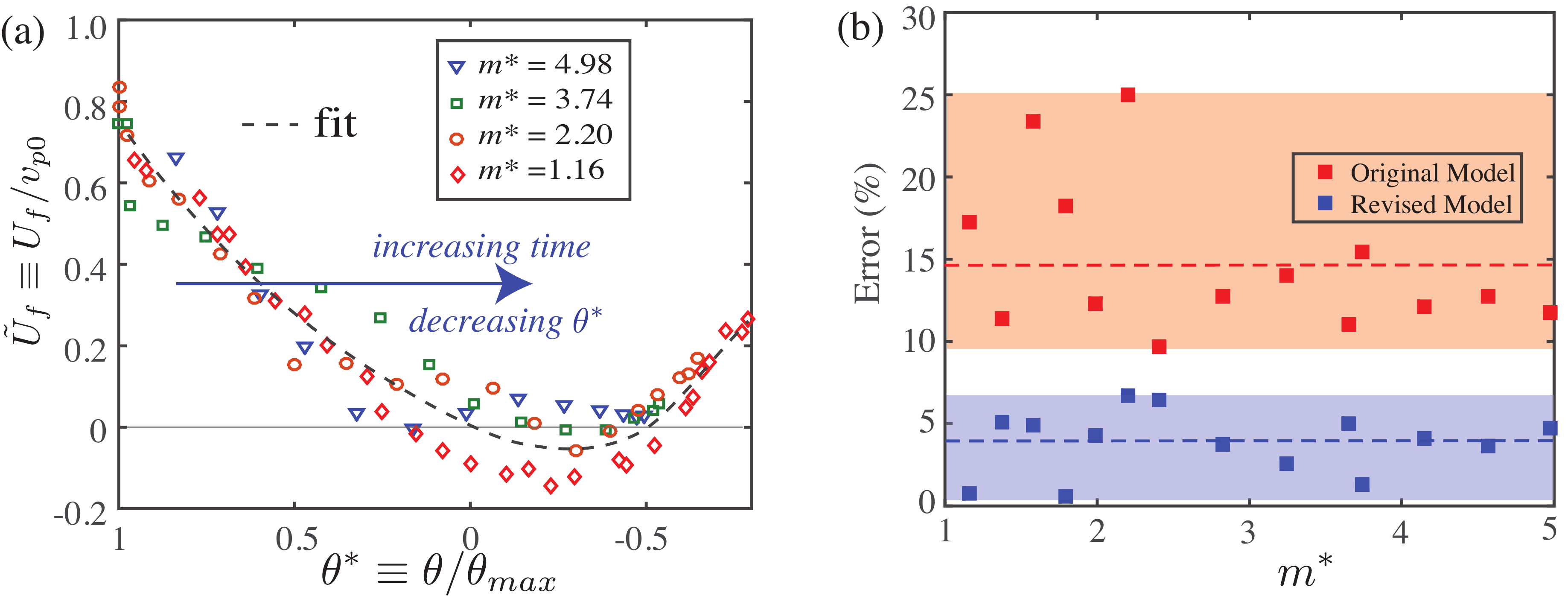}
	\caption{\textcolor{black}{(a)~Decay of the normalised wake velocity $\tilde{U}_f$ vs normalised angular position $\theta^*$. The wake velocity is normalised by the pendulum velocity at equilibrium position during the first swing $v_{p0}$, and the angular position is normalised by the peak angular position at the end of the first swing $\theta_{max}$. Both $v_{p0}$ and $\theta_{max}$ can be found by solving eq.~\ref{EqtofMotion}, and do not require any input from PIV. The normalisation leads to a reasonable collapse of the data despite the wide variation in Reynolds number from Re~$\sim$~4000~(for $m^* =1.16$) to  Re~$\sim$~34000~(for $m^* =4.98$). Note that the arrow of time is pointing opposite to the arrow of $\theta$, as indicated in the (a). (b)~Error (\%) in the peak amplitude after including the history force due to wake flow. The coloured bands show the error ranges for the original and revised model. The dashed lines show the mean errors for the two models, which decreases from 14.8\%~(for the original model) to 3.8\%~(for the revised model).}}
	\label{Wakedecay_thetanorm}
	\vspace{-.0cm}
\end{figure}

\vspace{-.5 cm}

\section{Conclusions and Discussion}

In this work, we have studied  the large amplitude oscillations of heavy and buoyant cylindrical pendulums in water.  The oscillation frequency and peak deceleration of the pendulum depend on the mass-density ratio $m^*$. Accounting for buoyancy and added mass alone, as done in previous investigations~\citep{neill2007pendulum}, does not  provide a complete picture of the dynamics.  We have developed a basic theoretical model, which accounts for  the nonlinear drag force and the bearing friction in addition to the buoyancy and the added mass. 
With the inclusion of these terms, the model predictions are in reasonable agreement with the experimental observations for a wide range of mass ratios. The added mass coefficient in experiment is found to be $m_a^* = 0.53$, i.e. lower that the two-dimensional~(2-D) potential flow value for a cylinder. \textcolor{black}{A reduced $m_a^*$ as compared to the 2-D potential flow added mass can be expected, given that the flow is viscous and three-dimensional, and also due to the finite span of the pendulum.  This result is consistent with existing literature on oscillating bodies in viscous fluids~\citep{dong1978effective,konstantinidis2013added,koo2015simplified}.}
The theoretical model presented here provides fair predictions of the oscillation frequency and peak decelerations for a wide range of $m^*$. However, the model overpredicts the peak oscillation amplitudes for the second and subsequent oscillations.  \textcolor{black}{Introducing a Reynolds number dependent drag, along with vortex shedding forces, can provide marginal improvements to the model predictions. While modelling $C_D$ as a function of Re improves the later oscillations, the vortex shedding term mainly influences the first oscillation.}

\textcolor{black}{Particle image velocimetry~(PIV) measurements have revealed that the major factor causing the deviations between experiment and the simplified model is the disturbed flow surrounding the cylinder, which was not accounted for in the original model. We have used insights from PIV measurements to obtain a simple model for the incoming wake flow for a wide range of mass ratios. With this history force modelled, the revised model predictions are significantly improved; the mean error reducing from 14.8 to 3.8\%. The predictions are also improved for other mass ratios for which PIV experiments were not performed.} 

\textcolor{black}{Even with the wake history force included, the current model is still quite basic. In reality, the dynamics is highly nonlinear, with changes in direction, curvilinear trajectories and wide variations in instantaneous Re. Under such conditions, exact analytical expressions for the drag, added mass and history forces are not available.
Fully resolved direct numerical simulations~(Immersed boundary~\citep{mittal2005immersed}, or Physalis~\citep{naso2010interaction}) can provide better insights into the flow-induced forces. 
On a different note, the present experiments have shown the importance of fluid drag for a bluff body oscillating in a fluid. An interesting extension to the study would be to use a more streamlined body, for which the nonlinear drag would be less dominant. In the absence of major flow separations, we can expect the frequency and dynamics to conform slightly better with a potential flow based added mass.}

We thank D. Lohse, C. Sun, A. Pandey and S. Maheshwari for fruitful discussions. We thank G.-W. Bruggert, M. Bos and D. van Gils for technical support. 
V.M. acknowledges funding from the Netherlands Organisation for Scientific Research NWO, and European Cooperation in Science and Technology, Eu-COST action MP1305.

\vspace{- .5 cm}
\bibliographystyle{jfm}
\bibliography{literatur_me_old.bib}

\end{document}